# Dirac particle in a square well and in a box

A. D. Alhaidari

*Saudi Center for Theoretical Physics, Dhahran, Saudi Arabia*

We obtain an exact solution of the 1D Dirac equation for a square well potential of depth greater then twice the particle's mass. The energy spectrum formula in the Klein zone is surprisingly very simple and independent of the depth of the well. This implies that the same solution is also valid for the potential box (infinitely deep well). In the nonrelativistic limit, the well-known energy spectrum of a particle in a box is obtained. We also provide in tabular form the elements of the complete solution space of the problem for all energies.

Aside from the mathematically "trivial" free case, particle in a box is usually the first problem that an undergraduate student of nonrelativistic quantum mechanics is asked to solve. Normally, he goes further into obtaining the bound states solution of a particle in a finite square well. It is much later that he works out more involved exercises like the 3D Coulomb problem [1]. Unfortunately, in relativistic quantum mechanics the story is reversed. One can hardly find a textbook on relativistic quantum mechanics where the 1D problem of a particle in a potential box is solved before the relativistic Hydrogen atom is [2]. The difficulty is that if this is to be done from the start, then one is forced to get into subtle issues like the Klein paradox, electron-positron pair production, stability of the vacuum, appropriate boundary conditions, …etc. [3]. In fact, the subtleties are so exasperating to the extent that Coulter and Adler ruled out this problem altogether from relativistic physics: "*This rules out any consideration of an infinite square well in the relativistic theory*" [4]. Moreover, in almost all earlier attempts at a solution of this problem, the bound states energy spectrum is obtained only approximately (graphically and/or numerically) [2,5,6]. To understand these difficulties in simpler terms, let us take the Schrödinger equation and the Dirac equation to describe the nonrelativistic and relativistic dynamics, respectively. Being a second order differential equation, the solution space of the former is much larger than that of the latter. This means that one will most likely succeed in finding a subspace of the solution space that satisfies the physical boundary conditions (i.e., finding a proper domain of the Hamiltonian). This makes the nonrelativistic Hamiltonian self-adjoint (i.e., has a real energy spectrum). However, in the relativistic case, the solution space of some problems is too small to support a self-adjoint structure for the Dirac Hamiltonian. Nevertheless, some attempts to overcome this problem were made by performing what is called "self-adjoint extension" of the Hamiltonian resulting in a larger solution space. For example, boundary conditions could be relaxed without violating the physics of the problem and/or the potential be slightly modified or regularized …etc. For the 1D Dirac equation with the box potential, this approach made it possible to find a subspace that satisfies appropriate boundary conditions in which a real energy spectrum could be obtained [7]. However, in earlier attempts to solve this problem, part of the solution of the Dirac equation is not accounted for. It is only when the missing part is included will we be able to obtain the correct analytic solution of the problem in a satisfactory and conventional manner.

In this article, we show that by accounting for the full contribution of the complete solution space of the 1D Dirac equation with the square well potential, a proper analytic solution is obtained without altering the physical configuration. We should emphasize that the potential enters in the Dirac Hamiltonian as the time component of a vector with vanishing space component. Inserting the same potential in the Dirac Hamiltonian as a scalar or admix of scalar and vector result in a physically different problem[†] which is less complicated [6]. The same approach that we will use here has recently been utilized in obtaining a resolution of the 80-years old Klein paradox [8]. We will show that by including all elements of the solution space, the exact relativistic energy spectrum and corresponding eigenfunctions are obtained. Moreover, the well-known nonrelativistic solution is easily found in the limit. A surprising result is that the relativistic energy spectrum is independent of the depth of the potential well as long as it is larger than twice the rest mass of the spinor. Before starting to solve the square well problem, we discuss briefly the square barrier. Figure 1 shows the physical configuration associated with the latter problem. The barrier height $V$ is greater than $2m$, where $m$ is the particle's mass, and the energy is in the range $+m < E < V - m$ (i.e., in the Klein energy zone). The boundary conditions are imposed by the physics of the problem. For example, a beam of free electrons is incident from left with a normalized unit amplitude gets partially reflected with an amplitude $R(E)$ and partially transmitted with amplitude $T(E)$. Reality gives: $|T|^2 + |R|^2 = 1$. Now, Fig. 2 shows the physical configuration associated with the square well problem. As far as the Dirac equation is concerned, this problem is a mirror image of the former. That is, the solution of the latter is obtained from the former simply by the replacement $E \to -E$ and $V \to -V$. However, a solution of a given differential equation is governed not only by the equation itself but also necessarily by the boundary conditions. The physics of the square well is not the same as that of the square barrier. In the latter only left or right incidence is possible at one time, whereas in the former both are allowed at the same time; and since the Dirac equation is linear, then a complete solution of the square well problem must be a linear combination of both. The concept of including the contribution of all possible processes before applying the physical constraints on the result is very common in physics. For example, in the path integral formulation of quantum mechanics, contributions from all possible trajectories (almost none is the physical path that the particle would eventually take) are included. It is only by imposing the physical constraint of minimal energy that the correct path is obtained. Another example is found in the off-shell calculation in quantum field theory where one includes the contribution from all possible momentum configurations, even those that violate $E^2 - \vec{p}^2 = m^2$. For the present problem and in addition to the continuity of the wavefunction across the walls of the square well, we impose the condition of a bounded state on the complete solution. This condition could be stated in different ways. For example,
1) The probability current across the walls of the well must vanish, or
2) The left particle flux inside the well must be balanced by the right flux (i.e., in Fig. 2, $|A|^2 = |B|^2$).

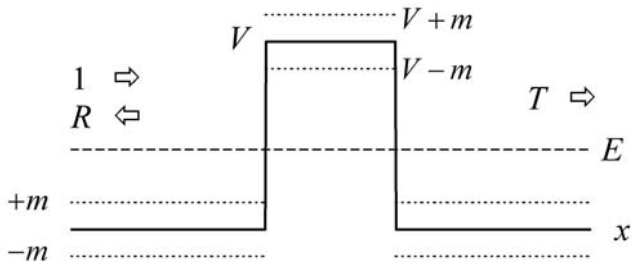
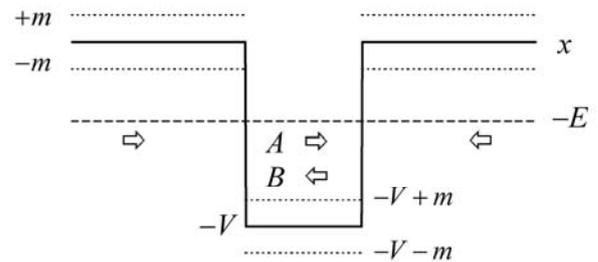

**Fig. 1**: Physical configuration of a square barrier potential with $V > 2m$. A beam of particles with coherent energy in the range $[+m, V-m]$ incident from left is partially reflected and transmitted.

**Fig. 2**: Physical configuration of a square well potential with $V > 2m$. Particles inside the well are maintained in a bound state at special discrete energies within the range $[-m, -V+m]$ by the combined effect of right and left incidence of negative energy anti-particle beams on either side of the well.

---

[†] A vector potential couples to the charge and spin. Thus, it interacts with anti-particles differently that with particles. However, a scalar potential couples to the mass. Therefore, it makes no distinction between the two. Another interpretation of scalar coupling is that the mass of the particle becomes position-dependent acquiring different values inside and outside the well.

These two conditions will also guarantee lepton/anti-lepton number conservation inside/outside the well. Next, we give a brief technical presentation of how to obtain the complete solution of the Dirac particle in a square well and in a box.

In the relativistic units $\hbar = c = 1$, the steady state Dirac equation for this 1D problem could be written as follows [9]

$$\begin{pmatrix} m + V(x) - E & -\frac{d}{dx} \\ \frac{d}{dx} & -m + V(x) - E \end{pmatrix} \begin{pmatrix} \psi^+(x) \\ \psi^-(x) \end{pmatrix} = 0. \tag{1}$$

The potential enters in the equation as the time component of a vector with vanishing space component. Outside the square well, where $V(x) = 0$, this equation relates the two spinor components as follows

$$\psi^{\mp}(x) = \frac{1}{m \pm E} \frac{d}{dx} \psi^{\pm}(x), \tag{2}$$

which is not valid for $E = \mp m$. We also obtain the following Schrödinger-like second order differential equation

$$\left( \frac{d}{dx^2} + E^2 - m^2 \right) \psi^{\pm}(x) = 0. \tag{3}$$

Since $E = \mp m$ belongs to the $\mp$tive energy spectrum, then Eq. (2) and Eq. (3) with the top/bottom sign are valid only for positive/negative energy, respectively. We should emphasize that Eq. (3) does not give the two components of the spinor that belong to the same energy subspace. One has to choose one sign in Eq. (3) to obtain only one of the two components then substitute that into Eq. (2) with the corresponding sign to obtain the other component. The positive and negative energy subspaces are completely disconnected. This is a general feature of the solution space of the Dirac equation, which is frequently overlooked. Now, inside the well, the same analysis follows but with the substitution $E \to E + V$.

We begin by giving the solution of the Dirac equation (1) for the square well located at $x \in [0, a]$ with a beam of negative energy anti-electron incident from *left*. The value of the energy is in the interval $-m > E > -V + m$. It is straightforward to write down the negative energy solution outside the potential well as

$$\psi(x) = \frac{1}{\sqrt{1+\alpha^2}} \begin{pmatrix} -i\alpha \\ 1 \end{pmatrix} e^{-ikx} + \frac{R}{\sqrt{1+\alpha^2}} \begin{pmatrix} i\alpha \\ 1 \end{pmatrix} e^{ikx}; \; x \leq 0, \tag{4a}$$

$$\psi(x) = \frac{T}{\sqrt{1+\alpha^2}} \begin{pmatrix} -i\alpha \\ 1 \end{pmatrix} e^{-ikx}; \qquad x \geq a, \qquad (4b)$$

where $k = \sqrt{E^2 - m^2}$ and $\alpha = \sqrt{(E+m)/(E-m)}$. Inside the potential well, the positive energy solution is

$$\psi(x) = \frac{A}{\sqrt{1+\beta^2}} \begin{pmatrix} 1 \\ i\beta \end{pmatrix} e^{ipx} + \frac{B}{\sqrt{1+\beta^2}} \begin{pmatrix} 1 \\ -i\beta \end{pmatrix} e^{-ipx}; \qquad 0 \leq x \leq a, \qquad (5)$$

where $p = \sqrt{(E+V)^2 - m^2}$ and $\beta = \sqrt{(E+V-m)/(E+V+m)}$. One should note that for positive (negative) energy, $e^{\pm iqx}$ is a wave traveling in the $\pm x$ ($\mp x$) direction, respectively, where $q$ is the positive wave number or linear momentum. Here, we followed the standard convention by normalizing the negative energy incident flux to unit amplitude. Physically, however, this amplitude is arbitrary whereas the electron current in the well is not. It is dictated by the experiment and, as such, could be normalized at will (e.g., by taking $|A|^2 + |B|^2 =$ constant). Nevertheless, one can show that the same solution is obtained either way. Now, matching the spinor wavefunction at $x = 0$ and $x = a$ gives

$$A = \frac{2i\alpha}{\alpha\beta+1} \sqrt{\frac{1+\beta^2}{1+\alpha^2}} \left[ e^{2ipa} \left( \frac{\alpha\beta-1}{\alpha\beta+1} \right)^2 - 1 \right]^{-1}, \quad B = \frac{\alpha\beta-1}{\alpha\beta+1} e^{2ipa} A. \qquad (6)$$

Repeating the same calculation but for a negative energy beam of anti-electron incident from *right*, we obtain

$$\hat{B} = \frac{-2i\alpha}{\alpha\beta+1} \sqrt{\frac{1+\beta^2}{1+\alpha^2}} e^{i(k+p)a} \left[ e^{2ipa} \left( \frac{\alpha\beta-1}{\alpha\beta+1} \right)^2 - 1 \right]^{-1}, \quad \hat{A} = \frac{\alpha\beta-1}{\alpha\beta+1} \hat{B}. \qquad (7)$$

The caret symbol on top refers to incidence from right. Consequently, the total spinor wavefunction inside the square well becomes

$$\Psi(x) = \psi(x) + \hat{\psi}(x) = \frac{\mathbb{A}}{\sqrt{1+\beta^2}} \begin{pmatrix} 1 \\ i\beta \end{pmatrix} e^{ipx} + \frac{\mathbb{B}}{\sqrt{1+\beta^2}} \begin{pmatrix} 1 \\ -i\beta \end{pmatrix} e^{-ipx}; \qquad 0 \leq x \leq a, \qquad (8)$$

where $\mathbb{A} = A + \hat{A}$ and $\mathbb{B} = B + \hat{B}$. Therefore, the electron flux inside the well is $|\mathbb{A}|^2$ to the right and $|\mathbb{B}|^2$ to the left. These are equal only at special values of the energy; the bound states energies. On the other hand, the current density inside the well is given by $J(x) = -i\Psi^\dagger \sigma_3 \sigma_1 \Psi$, where $\sigma_3 = \begin{pmatrix} +1 & 0 \\ 0 & -1 \end{pmatrix}$ and $\sigma_1 = \begin{pmatrix} 0 & 1 \\ 1 & 0 \end{pmatrix}$. Thus, we also expect that for bound state energies this current must vanish at the walls of the well. That is, for bound states, we require that $|\mathbb{A}|^2 = |\mathbb{B}|^2$ and $J(0) = J(a) = 0$. In fact, both conditions give the same result,

$$\cos(p+k)a = \cos(p-k)a. \qquad (9)$$

The two independent solutions of this equation are very simple. They are: $ka = n\pi$ and $pa = n\pi$, giving the following respective energy spectra

$$E_n = -m\sqrt{1 + (n\pi/ma)^2}, \text{ and} \qquad (10a)$$

$$E_n = -V + m\sqrt{1 + (n\pi/ma)^2}, \qquad (10b)$$

where $n = 0, 1, 2, ..., n_{max}$ and $n_{max}$ is the largest integer that is less than or equal to $\frac{ma}{\pi} \sqrt{\left(\frac{V}{m} - 1\right)^2 - 1}$.

The nonrelativistic limit is obtained by taking $ma \gg 1$ and $\frac{V}{m} \gg 1$. In Eq. (10a), this gives the correct well know energy spectrum of a particle in a box [1], $E_n^{NR} = n^2\pi^2/2ma^2$. Due to the interference of the right and left particle fluxes inside the potential well, the two conditions for obtaining the energy spectrum, $|\mathbb{A}|^2 = |\mathbb{B}|^2$ and $J(0) = J(a) = 0$, are week. It turns out that the necessary and sufficient condition, which is a subset of these two, is stronger and results uniquely in the spectrum formula (10a) while eliminating (10b). It belongs to the class of boundary conditions considered by Alonso and De Vincenzo [7] and reads as follows:

$$\begin{pmatrix} \Psi^+(a) \\ \Psi^-(a) \end{pmatrix} = \begin{pmatrix} \pm\Psi^+(0) \\ \mp\Psi^-(0) \end{pmatrix}. \tag{11}$$

That is $\Psi(a) = \pm\sigma_3\Psi(0)$, where the choice of sign depends to the parity of the eigenstate[‡]. For a state of energy $E_n$ this parity is $(-1)^{n+1}$. The energy spectrum formula (10a) is surprising, not only because it is so simple but also, because it states that the energy spectrum is independent of the depth of the square well, $V$, as long as $V > 2m$ and $-m > E > -V + m$ (i.e., if the energy is in the Klein zone). Consequently, the solution obtained above is also valid for the potential box where $\frac{V}{m} \gg 1$.

For completeness, we conclude by giving the full space of solution of the problem posed by Fig. 2 for all energies. This is shown in Table 1. The spinor basis elements for the complete solution space in the Table are defined as follows:

$$\left.\begin{array}{l} x > a \\ x < 0 \end{array}\right\} : \quad \phi_{\pm\uparrow}^-(x) = \frac{1}{\sqrt{1+\alpha^2}}\begin{pmatrix} 1 \\ \pm i\alpha \end{pmatrix}e^{\pm ikx}, \quad \phi_{\pm\downarrow}^-(x) = \frac{1}{\sqrt{1+\alpha^2}}\begin{pmatrix} \pm i\alpha \\ 1 \end{pmatrix}e^{\pm ikx} \tag{12}$$

$$\theta_{\pm\uparrow}^-(x) = \frac{1}{\sqrt{1+\alpha^2}}\begin{pmatrix} 1 \\ \pm\alpha \end{pmatrix}e^{\pm kx}, \quad \theta_{\pm\downarrow}^-(x) = \frac{1}{\sqrt{1+\alpha^2}}\begin{pmatrix} \pm\alpha \\ 1 \end{pmatrix}e^{\pm kx}$$

$$0 \leq x \leq a : \quad \phi_{\pm\uparrow}^+(x) = \frac{1}{\sqrt{1+\beta^2}}\begin{pmatrix} 1 \\ \pm i\beta \end{pmatrix}e^{\pm ipx}, \quad \phi_{\pm\downarrow}^+(x) = \frac{1}{\sqrt{1+\beta^2}}\begin{pmatrix} \pm i\beta \\ 1 \end{pmatrix}e^{\pm ipx} \tag{13}$$

$$\theta_{\pm\uparrow}^+(x) = \frac{1}{\sqrt{1+\beta^2}}\begin{pmatrix} 1 \\ \pm\beta \end{pmatrix}e^{\pm px}, \quad \theta_{\pm\downarrow}^+(x) = \frac{1}{\sqrt{1+\beta^2}}\begin{pmatrix} \pm\beta \\ 1 \end{pmatrix}e^{\pm px}$$

The energy dependent quantities in these spinor functions are

$$k = \sqrt{|E^2 - m^2|}, \quad p = \sqrt{|(E+V)^2 - m^2|}, \quad \alpha = \sqrt{\left|\frac{|E|-m}{|E|+m}\right|}, \quad \beta = \sqrt{\left|\frac{|E+V|-m}{|E+V|+m}\right|}. \tag{14}$$

The + and − superscript sign refers to the inside and outside of the potential well, respectively. The meaning of the sign in the subscript is obvious. The up/down arrow in the subscript signifies positive/negative energy solution.

---

[‡] A condition other than (11) gives uniquely the energy spectrum (10b) and eliminates the desirable spectrum formula (10a). It is also a subset of the particles flux balance and current quenching conditions and it reads as follows: $\Psi(a) = \pm\Psi(0)$.

**Table 1:** The full solution space of the problem portrayed in Fig. 2 for all energies. The two-component spinor basis functions are defined by equations (12-14).

| Energy range | $\psi(x<0)$ | $\psi(0 \leq x \leq a)$ | $\psi(x>a)$ |
|---|---|---|---|
| $E > +m$ | $\phi^-_{+\uparrow}(x) + R\,\phi^-_{-\uparrow}(x)$ | $A\,\phi^+_{+\uparrow}(x) + B\,\phi^+_{-\uparrow}(x)$ | $T\,\phi^-_{+\uparrow}(x)$ |
| $+m > E > 0$ | $C\,\theta^-_{+\uparrow}(x)$ | $A\,\phi^+_{+\uparrow}(x) + B\,\phi^+_{-\uparrow}(x)$ | $D\,\theta^-_{-\uparrow}(x)$ |
| $0 > E > -m$ | $C\,\theta^-_{+\downarrow}(x)$ | $A\,\phi^+_{+\uparrow}(x) + B\,\phi^+_{-\uparrow}(x)$ | $D\,\theta^-_{-\downarrow}(x)$ |
| $-m > E > -V+m$ | $\phi^-_{-\downarrow}(x) + R\,\phi^-_{+\downarrow}(x)$ $\hat{T}\,\phi^-_{+\downarrow}(x)$ | $\mathbb{A}\,\phi^+_{+\uparrow}(x) + \mathbb{B}\,\phi^+_{-\uparrow}(x)$ | $T\,\phi^-_{-\downarrow}(x)$ $\phi^-_{+\downarrow}(x) + \hat{R}\,\phi^-_{-\downarrow}(x)$ |
| $-V+m > E > -V$ | $\phi^-_{-\downarrow}(x) + R\,\phi^-_{+\downarrow}(x)$ $\hat{T}\,\phi^-_{+\downarrow}(x)$ | $\mathbb{A}\,\theta^+_{+\uparrow}(x) + \mathbb{B}\,\theta^+_{-\uparrow}(x)$ | $T\,\phi^-_{-\downarrow}(x)$ $\phi^-_{+\downarrow}(x) + \hat{R}\,\phi^-_{-\downarrow}(x)$ |
| $-V > E > -V-m$ | $\phi^-_{-\downarrow}(x) + R\,\phi^-_{+\downarrow}(x)$ $\hat{T}\,\phi^-_{+\downarrow}(x)$ | $\mathbb{A}\,\theta^+_{+\downarrow}(x) + \mathbb{B}\,\theta^+_{-\downarrow}(x)$ | $T\,\phi^-_{-\downarrow}(x)$ $\phi^-_{+\downarrow}(x) + \hat{R}\,\phi^-_{-\downarrow}(x)$ |
| $E < -V-m$ | $\phi^-_{-\downarrow}(x) + R\,\phi^-_{+\downarrow}(x)$ $\hat{T}\,\phi^-_{+\downarrow}(x)$ | $\mathbb{A}\,\phi^+_{-\downarrow}(x) + \mathbb{B}\,\phi^+_{+\downarrow}(x)$ | $T\,\phi^-_{-\downarrow}(x)$ $\phi^-_{+\downarrow}(x) + \hat{R}\,\phi^-_{-\downarrow}(x)$ |